\documentclass{osa-article}

\journal{oe}

\articletype{Research Article}

\usepackage[percent]{overpic}
\usepackage{upgreek}

\newcommand{\mum}{$\upmu$m }
\newcommand{\mumN}{$\upmu$m}
\newcommand{\mus}{$\upmu$s }
\newcommand{\musN}{$\upmu$s}
\newcommand{\muj}{$\upmu$J }
\newcommand{\mujN}{$\upmu$J}

\usepackage[overlay,absolute]{textpos}
\newcommand\PlaceText[3]{%
	\begin{textblock*}{10in}(#1,#2)
		#3
	\end{textblock*}
}%

\begin{document}

\title{Q-switched Dy:ZBLAN fiber lasers beyond 3~$\upmu$m: comparison of pulse generation using acousto-optic modulation and inkjet-printed black phosphorus}

\author{R. I. Woodward\authormark{1}, M. R. Majewski\authormark{1}, N.~Macadam\authormark{2}, G.~Hu\authormark{2}, T.~Albrow-Owen\authormark{2}, T.~Hasan\authormark{2, *} and S. D. Jackson\authormark{1}}

\address{\authormark{1} MQ Photonics, School of Engineering, Macquarie University, New South Wales, Australia}
\address{\authormark{2} Cambridge Graphene Centre, University of Cambridge, Cambridge, UK}

\email{\authormark{*}th270@cam.ac.uk} 

\begin{abstract}
We report high-energy mid-infrared pulse generation by Q-switching of dysprosium-doped fiber lasers for the first time.
Two different modulation techniques are demonstrated.
Firstly, using active acousto-optic modulation, pulses are produced with up to 12~\muj energy and durations as short as 270~ns, with variable repetition rates from 100~Hz to 20~kHz and central wavelengths tunable from 2.97 to 3.23~\mumN.
Experiments are supported by numerical modeling, identifying routes for improved pulse energies and to avoid multi-pulsing by careful choice of modulator parameters.
Secondly, we demonstrate passive Q-switching by fabricating an inkjet-printed black phosphorus saturable absorber, simplifying the cavity and generating 1.0~\muj pulses with 740~ns duration.
The performance and relative merits of each modulation approach are then critically discussed.
These demonstrations highlight the potential of dysprosium as a versatile gain medium for high-performance pulsed sources beyond 3~\mumN.
\end{abstract}

\PlaceText{25mm}{9mm}{Optics Express, Vol. 27, Issue 10, pp. 15032--15045 (2019); https://doi.org/10.1364/OE.27.015032}

\section{Introduction}
Mid-infrared (mid-IR) fiber lasers are currently emerging as auspicious high-brightness sources of light beyond 2.5~\mumN.
While a number of laser architectures (e.g. nonlinear parametric wavelength conversion and bulk Cr/Fe-doped crystals) offer similar long-wavelength emission, the compact footprint and flexible nature of the fiber platform are practically advantageous, paving the way to deployable systems for real-world applications.

Indeed, it is the vast array of potential applications that is driving the current research interest in mid-IR source development, arising from the existence of strong rotational and vibrational molecular absorption lines in this spectral region.
For example, mid-IR fiber lasers in swept-wavelength operation have recently been demonstrated for real-time gas sensing, probing multiple mid-IR absorption lines simultaneously~\cite{Woodward2019_nh3}, in addition to being used for polymer ablation by targeting resonant C-H bond absorption~\cite{Frayssinous2018}.
In both these experiments, continuous wave (CW) light was used.
There is pressing demand, however, for high-energy \emph{pulsed} mid-IR sources, to enable time-resolved sensing (e.g. LIDAR) and improved laser machining throughput.

To generate such pulses, gain-switching and Q-switching techniques have been widely applied to both erbium- (Er) and holmium-doped (Ho) single-mode fluoride fiber lasers emitting in the 2.7 to 3.0~\mum region~\cite{Frerichs1996a, Wei2012b, Wei2013, Gorjan2011a, Hu2013b, Li2012e, Li2014l}.
Pulse energies up to 10s \muj have been achieved, with durations as short as $\sim$100~ns.
Demand exists for longer wavelengths, however, e.g. to exploit the 3--5~\mum atmospheric transparency window for LIDAR.
To meet this need, a dual-wavelength-pumped Er transition is currently being explored for 3.4--3.8~\mum emission, which has been Q-switched~\cite{Bawden2018, Qin2018} and gain-switched~\cite{Jobin2018}, generating up to 7.8~\muj pulses.
Despite these advances, there still exists a 3.0--3.4~\mum `gap' in current fiber laser coverage, highlighting a need to consider alternative rare-earth-doped fibers.

In recent years, dysprosium (Dy) has surfaced as an ideal ion for next-generation mid-IR fiber lasers.
Originating from a spectroscopically simple transition (i.e. not requiring dual-wavelength-pumping or energy transfer mechanisms) from the first excited state to the ground state, broadly tunable CW emission from 2.8 to 3.4~\mum has been demonstrated~\cite{Majewski2018} in addition to watt-level powers with high slope efficiency through in-band pumping~\cite{Woodward2018_watt}.
Although Dy picosecond mode-locking using a frequency-shifted feedback mechanism has recently been reported~\cite{Woodward2018_fsf}, in addition to a numerical study on the prospects for gain switching~\cite{Falconi2018a}, high-energy pulsed Dy lasers have not yet been demonstrated.

Here, we report Q-switching of Dy-doped fiber lasers for the first time, considering both active Q-switching using acousto-optic modulation, and passive Q-switching with a black phosphorus saturable absorber. 
Prospects for high-energy mid-IR pulse generation with dysprosium are discussed with respect to the ion's spectroscopy, in addition to considering optimum Q-switch designs, supported by numerical modeling.

\section{Active Q-switching using acousto-optic modulation}
\label{sec:active}
We first consider active modulation using an acousto-optic tunable filter (AOTF).
The AOTF comprises an anisotropic TeO$_2$ crystal in `slow-shear operation'~\cite{Chang1981, Ward2018}: the application of an RF (MHz) sinusoidal voltage to an attached transducer generates acoustic shear waves which propagate through the crystal.
When light is incident on the device, only a narrow band of optical wavelengths meet the phase-matching condition for constructive interference with the acoustic waves. 
These wavelengths are then diffracted at a different angle to the undiffracted light, so the device acts as a spectral filter.
The central wavelength is tunable by varying the MHz drive frequency, since this changes the frequency of the generated acoustic waves and thus, the phase-matching condition.
To operate the AOTF as a Q-switch, the MHz sinusoid signal is gated by a square wave modulation function, effectively pulsing the drive signal on and off and varying the ratio of incident light that is diffracted into the 1$^\mathrm{st}$ order beam (the remaining light is transmitted through the AOTF undiffracted, known as the 0$^\mathrm{th}$ order).
We note that the AOTF used in our experiments (Gooch \& Housego) has a 5~nm filter bandwidth, 25~\mus rise time and 75\% maximum diffraction efficiency.

\begin{figure}[tb]
	\centering
	\sffamily
	\begin{overpic}[width=\textwidth]{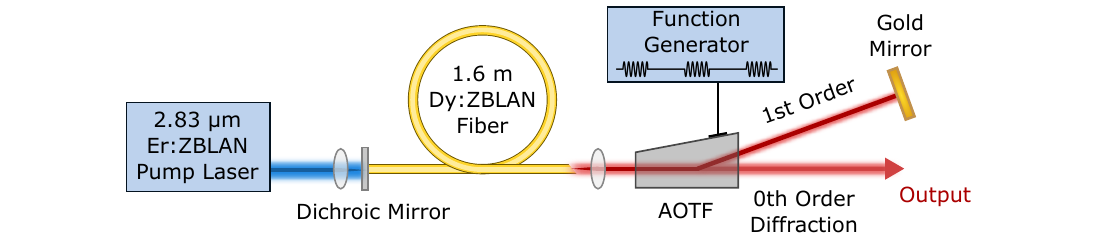}
	\end{overpic}
	\rmfamily
	\caption{Actively Q-switched Dy:ZBLAN laser schematic.}
	\label{fig:cavity}
\end{figure}

The AOTF is included in a linear cavity with 1.6~m of 2000 mol. ppm Dy:ZBLAN fiber (12.5~\mum core diameter, 0.16 NA, Le Verre Fluor\'{e}) and an input dichroic mirror which is broadly reflective beyond 2.95~\mum (Fig.~\ref{fig:cavity}).
The 0$^\mathrm{th}$-order (undiffracted light) from the AOTF is taken as the output (which passes through a filter to cut out any unabsorbed pump light) and the diffracted light is resonated by a broadband reflective gold mirror.

\subsection{Experimental results}
To demonstrate typical laser performance, the AOTF RF drive frequency is set to 18.1~MHz, corresponding to 3.1~\mum filter center wavelength.
This wavelength cannot be generated by the current generation of rare-earth-doped fibers, highlighting the benefits of the dysprosium ion.
Under these conditions, the threshold pump power is 400~mW.

For Q-switched pulse generation, square wave modulation is applied to the RF sinusoid in the function generator, with modulation frequency $f_\mathrm{mod}$ and duty cycle $D=\tau_\mathrm{ON}f_\mathrm{mod}$, where $\tau_\mathrm{ON}$ is the AOTF on-time per cycle (i.e. forming a high-Q cavity by diffracting light towards the gold mirror, so the intracavity field builds up by extracting gain from the excited Dy fiber).
For optimal operation, maximum energy should be extracted from the gain medium in a single pulse per modulation cycle, i.e. avoiding the generation of satellite pulses (`multi-pulsing'), which depend critically on the modulation parameters~\cite{Upadhyaya2011}.
Experimentally, we found that if the modulator is on for too short a time within each cycle, then lasing is suppressed, but if the on-time is too long (i.e. duty cycle too high), then multiple pulses are formed per cycle.
Therefore, for all experiments we empirically optimized the modulation signal to achieve single-pulse Q-switching.

With a pump power of 450~mW, we found that a constant AOTF on-time of $\tau_\mathrm{ON}=20$~\mus ensured stable single pulsing across a wide range of modulation frequencies from 100 Hz to 20~kHz. 
The duty cycle thus needed to be set accordingly for each modulation frequency.
As the pump power was increased, shorter AOTF on-times were required (16~\mus for 600~mW pump, 14.5~\mus for 750~mW pump) to avoid multi-pulsing (further discussion of the laser dynamics related to this phenomena, alongside simulations, are presented in Section~\ref{sec:numerics}).

\begin{figure}[tb]
	\centering
	\sffamily
	\begin{overpic}{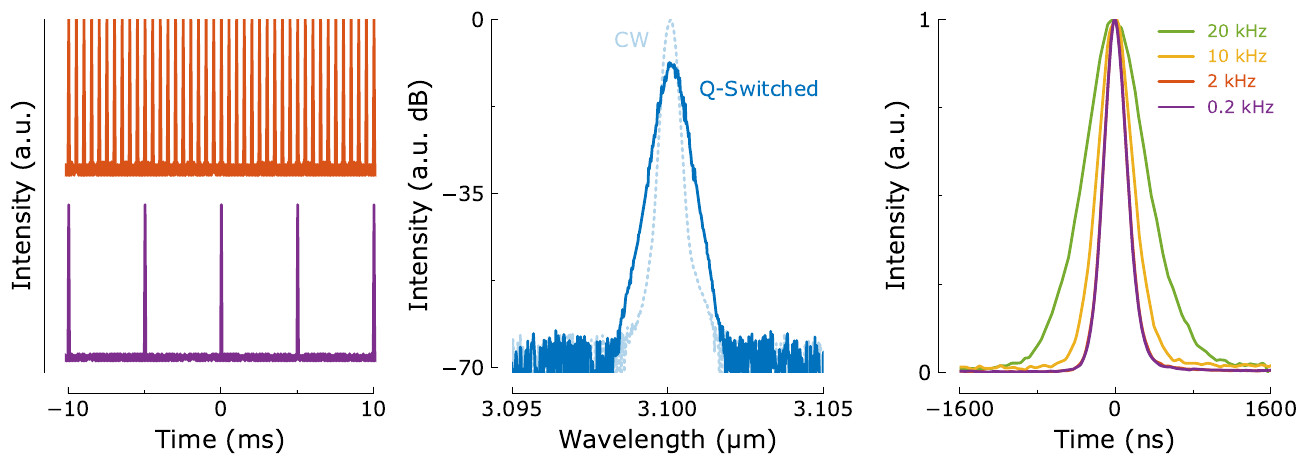}
		\put(-1, 34){ {\small (a)} }
		\put(29, 34){ {\small (b)} }
		\put(65, 34){ {\small (c)} }
	\end{overpic}
	\rmfamily
	\caption{Actively Q-switched laser characterization at 3.1~\mum with 450~mW pump power: (a) pulse trains at 0.2~kHz (bottom) and 2~kHz (top); (b) optical spectrum; (c) pulse shapes at 0.2~kHz, 2~kHz, 10~kHz and 20~kHz.}
	\label{fig:3100_pulses}
\end{figure}

With optimized parameters, the laser produced a stable Q-switched train of Gaussian-shaped pulses, at a repetition rate corresponding to the modulation frequency [Fig.~\ref{fig:3100_pulses}(a)]. 
A single peak at 3.1~\mum is observed in the optical spectrum [Fig.~\ref{fig:3100_pulses}(b)].
Compared to CW operation (i.e. with no AOTF modulation), the Q-switched laser spectrum is slightly broadened, likely due to self-phase modulation from increased peak power.
The pulse repetition rate could be continuously varied from 100~Hz to 20~kHz, maintaining stable pulsation, by adjusting the AOTF modulation frequency.
As the repetition rate was increased, the pulse duration was initially constant, but above a few kHz frequencies, significant broadening was observed [Fig.~\ref{fig:3100_pulses}(c)].

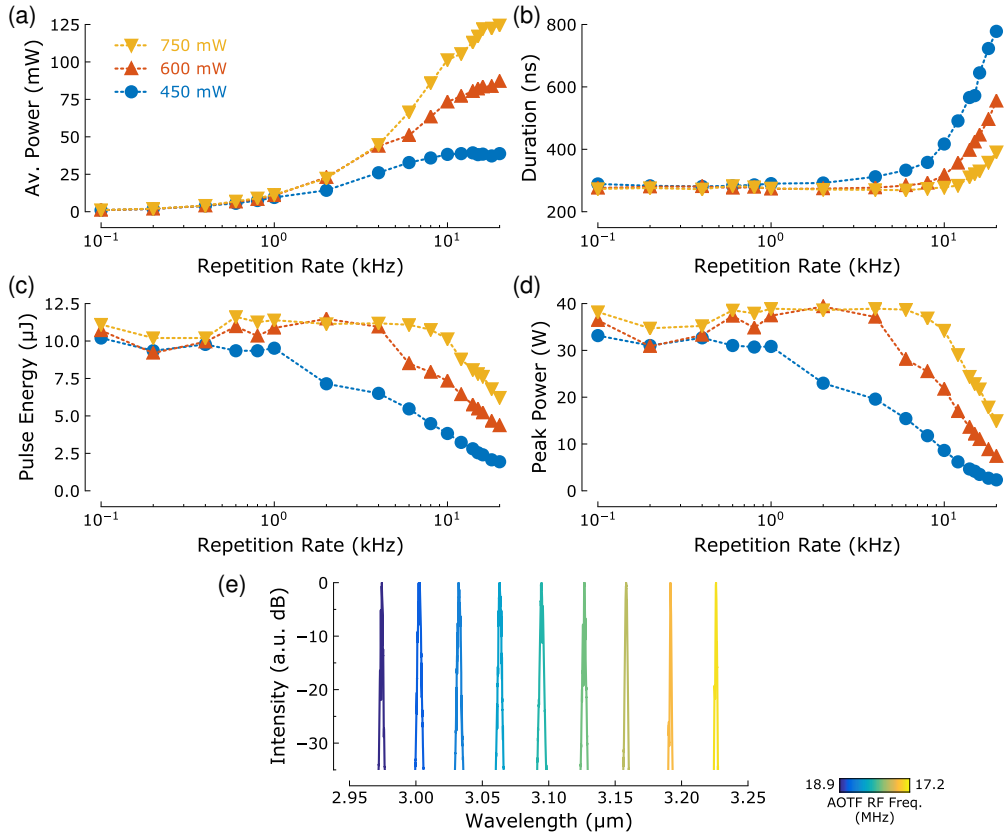
\begin{figure}[bt]
	\centering
	\sffamily
	\begin{overpic}{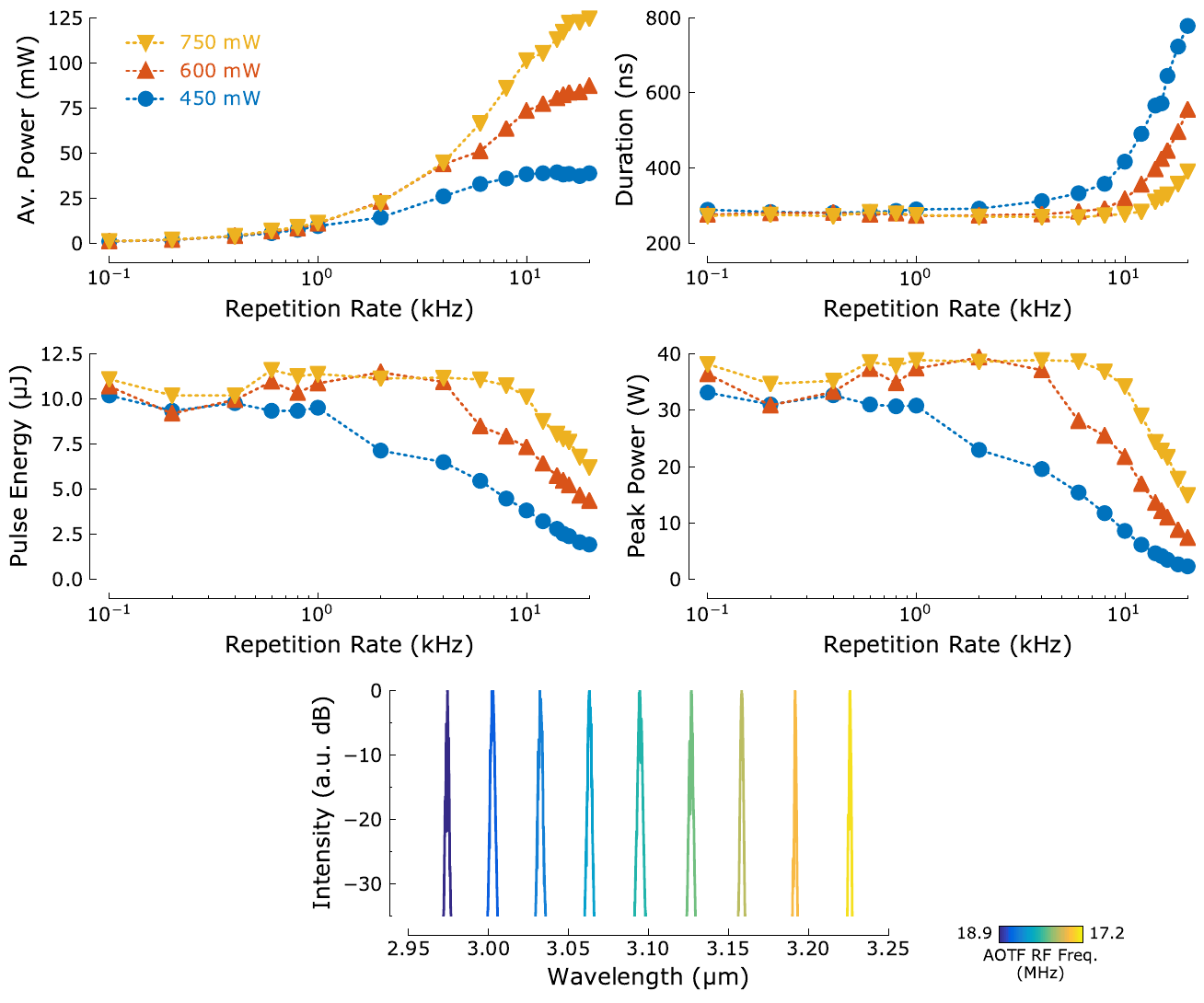}
		\put(-1, 82){ {\small (a)} }
		\put(49.5, 82){ {\small (b)} }
		\put(-1, 55){ {\small (c)} }
		\put(49.5, 55){ {\small (d)} }
		\put(20, 25){ {\small (e)} }
	\end{overpic}
	\rmfamily
	\caption{Variation of actively Q-switched 3.1~\mum laser pulse properties with respect to repetition rate: (a) average power; (b) duration; (c) pulse energy and (d) peak power. (e) Output spectra for Q-switched operation at various wavelengths within the tuning range.}
	\label{fig:3100_char}
\end{figure}

To understand the potential pulse parameter space offered by our laser, the output was fully characterized from 100~Hz to 20~kHz repetition rates for different pump powers (Fig.~\ref{fig:3100_char}).
The observed behavior is typical of repetitively Q-switched lasers and can be explained by considering two distinct regions for any given pump power~\cite{Siegman1986}.
At low repetition rates, pump absorption is saturated during the long time window when lasing is inhibited (i.e. when the AOTF is off)~\cite{Chesler1970}. 
Since the energy that can be extracted from an excited gain medium (and associated build-up dynamics) depends on how far above threshold it is pumped, this saturation results in near-constant pulse energies and pulse durations at low repetition rates.
Increasing the repetition rate linearly increases the average power, since more constant-energy pulses are emitted per second.
At high repetition rates (with shorter time for pump saturation effects), however, the average power approaches a constant value, approximately equal to output power if the laser were operated under CW conditions.
Increasing the pulse frequency here thus results in a reduction of the pulse energy.
Consequently, the pulse duration broadens since reduced pulse energies lead to weaker modulation of the net gain and thus, slower rise/decay of optical power.
Therefore, active Q-switching offers a near-constant peak-power enhancement factor at low repetition rates, but this falls sharply at frequencies above $\sim$1.5~kHz.

The transition between these two operating regions occurs when the repetition rate is on the order of the spontaneous excited population decay rate.
For dysprosium, with an upper-state lifetime of $\tau=650$~\musN~\cite{Gomes2010}, this transition point is expected to occur at approximately $f_\mathrm{rep}=1/\tau=$1.5~kHz, which is well supported by our experimental data (Fig.~\ref{fig:3100_char}).

Higher pump power leads to a greater average output power in the high repetition rate region, as expected, and the greater pump rate yields a slower change in duration and pulse energy with increasing repetition frequency, making high pump power favorable for optimum performance at higher repetition rates.
The maximum pulse energy, however, was not significantly changed, which may also be related to the need to reduce the AOTF on-time for higher pump powers in order to maintain stable single-pulse operation.
Summarizing this data, Fig.~\ref{fig:3100_char} shows that the laser can produce Q-switched pulses as short as $\tau=$270~ns, with pulse energies, $E$, up to 12~\mujN, corresponding to peak powers as large as 39~W (calculated using $P_\mathrm{peak}=0.94 E/\tau$ where 0.94 is the shape factor for Gaussian pulses).

Finally, we note that the AOTF permits tuning within the wide Dy gain bandwidth. 
By varying the applied RF sinusoid frequency from 17.2~MHz to 18.9~MHz, the laser could be tuned from 2.97~\mum to 3.23~\mumN.
Q-switching could be achieved at each wavelength within this range [Fig.~\ref{fig:3100_char}(e)], although the modulator on-time had to be re-optimized each time.
This is as expected, since the Dy emission cross-section varies with wavelength, which affects the gain and thus influences the inversion dynamics.
We note, however, that with the emergence of self-tuning laser designs, e.g. exploiting recent advances in machine intelligence~\cite{Woodward_ol_2017,Kutz2015}, it is foreseeable that the modulation parameter optimization process could be automated for on-demand output properties---particularly as the temporal and spectral filtering here are already entirely electronically controlled with no moving parts.
Additionally, it is expected that the tuning range could be increased by reducing cavity losses, as discussed in Ref.~\cite{Woodward2019_nh3}; for example, using a lower loss cavity with diffraction grating tuning, CW lasing with Dy fiber has been demonstrated from 2.8 to 3.4~\mum \cite{Majewski2018}.

\subsection{Numerical modeling methods}
\label{sec:numerics}
To gain insight into our laser dynamics and identify how to optimize the cavity for higher pulse energies, we developed a rate-equation-based numerical model, including measured cross sections and spectroscopic parameters~\cite{Gomes2010}. 
Fortunately, only the ground ($^6H_{15/2}$) and first excited level ($^6H_{13/2}$) of Dy need to be included in the simulation, (i.e. modeling a simple two-level system with ground state population $N_0$ and excited state  population $N_1$, where $N_0 + N_1 = N_\mathrm{total} = 2000$ mol. ppm$ = 3.66\times 10^{25}$ m$^{-3}$) for the case of in-band pumping, because there are no known excited state absorption (ESA) or energy transfer upconversion (ETU) transitions at these wavelengths~\cite{Gomes2010}.

The power evolution $P^+(\lambda, z)$ and $P^-(\lambda, z)$ along the Dy fiber (with longitudinal co-ordinate $z$) for each spectral channel of wavelength $\lambda$ in a given direction ($+$ forwards, $-$ backwards) is governed by:
\begin{equation}
\frac{\mathrm{d}P^\pm(\lambda, z)}{\mathrm{d}z} = \pm \left(
P^\pm(\lambda, z) \left[
g(\lambda, z) - l
\right] + P_\text{spon}(\lambda)
\right)
\end{equation}
where the gain is:
\begin{equation}
g(\lambda, z) = \Gamma(\lambda) \left[
\sigma_{10}(\lambda) N_1(z) - \sigma_{01}(\lambda) N_0(z)
\right]
\end{equation}
and where $\sigma_{10}(\lambda)$ is the emission cross section, $\sigma_{01}(\lambda)$ is the absorption cross section, $\Gamma$($\lambda$) is the overlap factor of the guided mode with the doped core and $l$ is the background loss (measured as 0.3~dB m$^{-1}$ at 3.39~\mum and assumed constant over the lasing range).

Spontaneous emission $P_\text{spon}$ is included to account for amplified spontaneous emission (ASE) effects using~\cite{Giles1991}:
\begin{equation}
P_\text{spon}(\lambda) = \Gamma \sigma_{10}(\lambda) N_1(z) E_\mathrm{ph} \Delta f.
\end{equation}
where $E_\mathrm{ph}=hc/\lambda$ is the photon energy and $\Delta f$ is the spectral width of the numerically defined channels.
While this spontaneous emission term can often be neglected for steady-state CW laser modeling (i.e. when $\frac{dN}{dt}$ set equal to 0 and solved as boundary value problem), it is important to include for pulsed modeling, since this seeds the amplification process which describes the repetitive pulse build-up and decay.
The $\Delta f$ parameter must be chosen carefully and different approaches currently exist in the literature.
The most accurate, yet computationally most expensive, approach is to define many wavelength channels with narrow spectral width (e.g. $\sim$1 nm) and numerically solve for all channels simultaneously~\cite{Giles1991, Huo2005}.
However, effective simplifications have been introduced such as using a single numerical channel for the laser wavelength with spectral width parameter $\Delta f$ set equal to either the gain bandwidth, or some fraction of this (accounting for gain narrowing / bandpass filtering effects)~\cite{Huo2005, Sujecki2018, Falconi2018a}.
Here, we use a single numerical channel for the signal (in addition to a single pump channel) with 5~nm spectral width, corresponding to the filter bandwidth of our AOTF; this approach is validated by the good agreement we observe with experiments (and with a reasonable computational time on the order of minutes, implemented using Python on a personal computer).

The atomic level populations $N_0$ and $N_1$ are governed by rate equations at each $z$ position:
\begin{multline}
\frac{\mathrm{d}N_1(t)}{\mathrm{d}t} = -N_1(t) \left( \frac{1}{\tau} + 
\sum_{\lambda} \frac{\sigma_{10}(\lambda) \Gamma(\lambda) [P^+(\lambda) + P^-(\lambda)]}{A_\text{core} \times h c / \lambda}
\right) 
+ \\
N_0(t) \left(
\sum_{\lambda} \frac{\sigma_{01}(\lambda) \Gamma(\lambda) [P^+(\lambda) + P^-(\lambda)]}{A_\text{core} \times h c / \lambda}
\right)
\end{multline}
where $\tau=650$~$\upmu$s is the upper state lifetime~\cite{Gomes2010}, $A_\text{core}$ is the geometric core area, $h$ is Planck's constant and $c$ is the speed of light. 
Note that Eqn.~4 includes the effect of all spectral channels through the summation term ($\sum_{\lambda}$): here we consider only 1 pump ($\lambda_p=2.825$~\mumN) and 1 signal ($\lambda_s=3.1$~\mumN) channel in each direction, although this approach is scalable to any number of spectral channels.
For convenience, Table~\ref{tab:model_params} summarizes our modeling parameters.

To efficiently solve these equations, we note that the 2 independent variables, $t$ and $z$, are coupled by the group velocity of light in the fiber, $n_\mathrm{g}$.
The fiber length is thus discretized into steps of length $\Delta L$ corresponding to a time step $\Delta t = \frac{\Delta L}{c / n_\mathrm{g}}$ where $n_\mathrm{g}\sim1.5$~\cite{Malouf2016}.
The atomic population and power evolution equations can then be integrated in time using a fixed-step 4th-order Runge-Kutta method, where the power values at each position are shifted one distance step along the fiber each time step.
At the fiber ends, a fraction of the signal power is taken as the output and the remainder is injected back into the fiber traveling in the opposite direction, thus accounting for mirror reflectivities.
Pump power is also injected at the input fiber end.

\begin{table}[htbp]
	\centering
	\caption{\bf Modeling parameters for the Dy:ZBLAN fiber laser}
	\begin{tabular}{ccc}
		\hline
		Parameter & Symbol & Value \\
		\hline
		Pump wavelength & $\lambda_p$ & 2.825~\mum \\
		Pump absorption cross section & $\sigma_{01}(\lambda_p)$ & 3.9 $\times 10 ^{-25}$ m$^{-3}$ \\
		Pump emission cross section & $\sigma_{10}(\lambda_p)$ & 2.6 $\times 10 ^{-25}$ m$^{-3}$ \\
		Pump overlap factor & $\Gamma(\lambda_p)$ & 0.79 \\
		Signal wavelength & $\lambda_s$ & 3.100~\mum \\
		Signal absorption cross section & $\sigma_{01}(\lambda_s)$ & 2.8 $\times 10 ^{-26}$ m$^{-3}$ \\
		Signal emission cross section & $\sigma_{10}(\lambda_s)$ & 6.8 $\times 10 ^{-26}$ m$^{-3}$ \\
		Signal overlap factor & $\Gamma(\lambda_s)$ & 0.75 \\
		Upper-state lifetime & $\tau$ & 650~\mus  \\
		Doping concentration & $N_\mathrm{total}$ & $3.66\times 10^{25}$ m$^{-3}$  \\
		Fiber background loss & $l$ & 0.3 dB m$^{-1}$ \\
		Numerical channel spectral width  & $\Delta f$ & 5~nm \\
		Core area & $A_\mathrm{core}$ & 123 $\upmu$m$^2$ \\
		\hline
	\end{tabular}
	\label{tab:model_params}
\end{table}

\subsection{Simulation results}
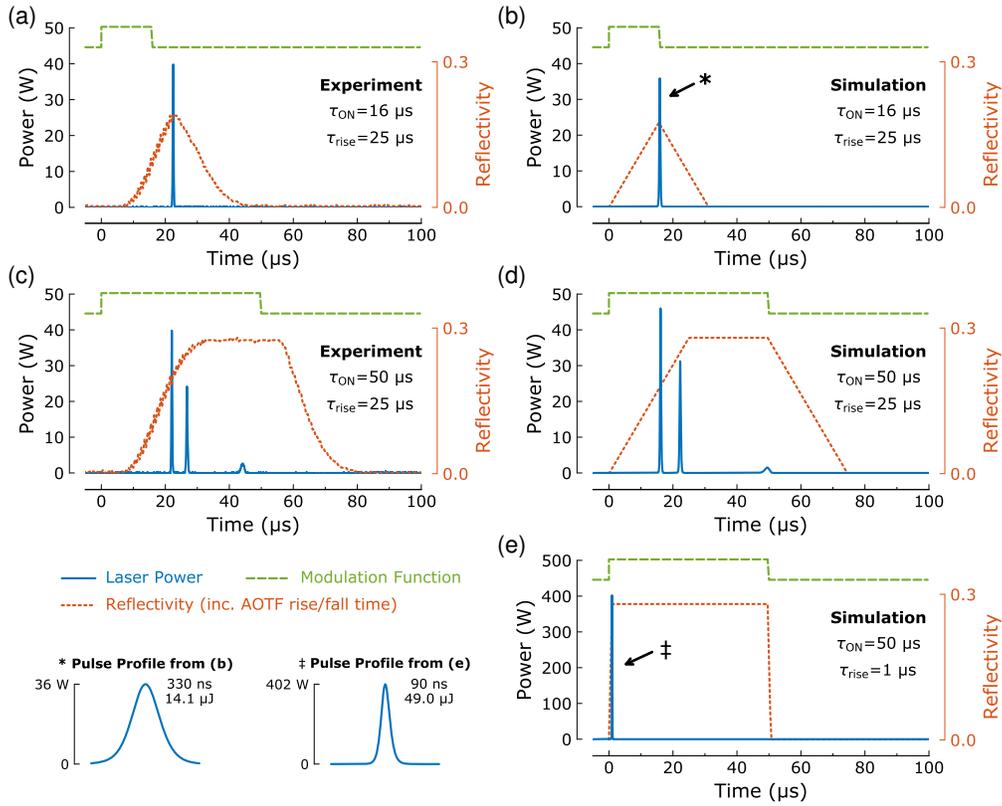
\begin{figure}[tb]
	\centering
	\sffamily
	\begin{overpic}{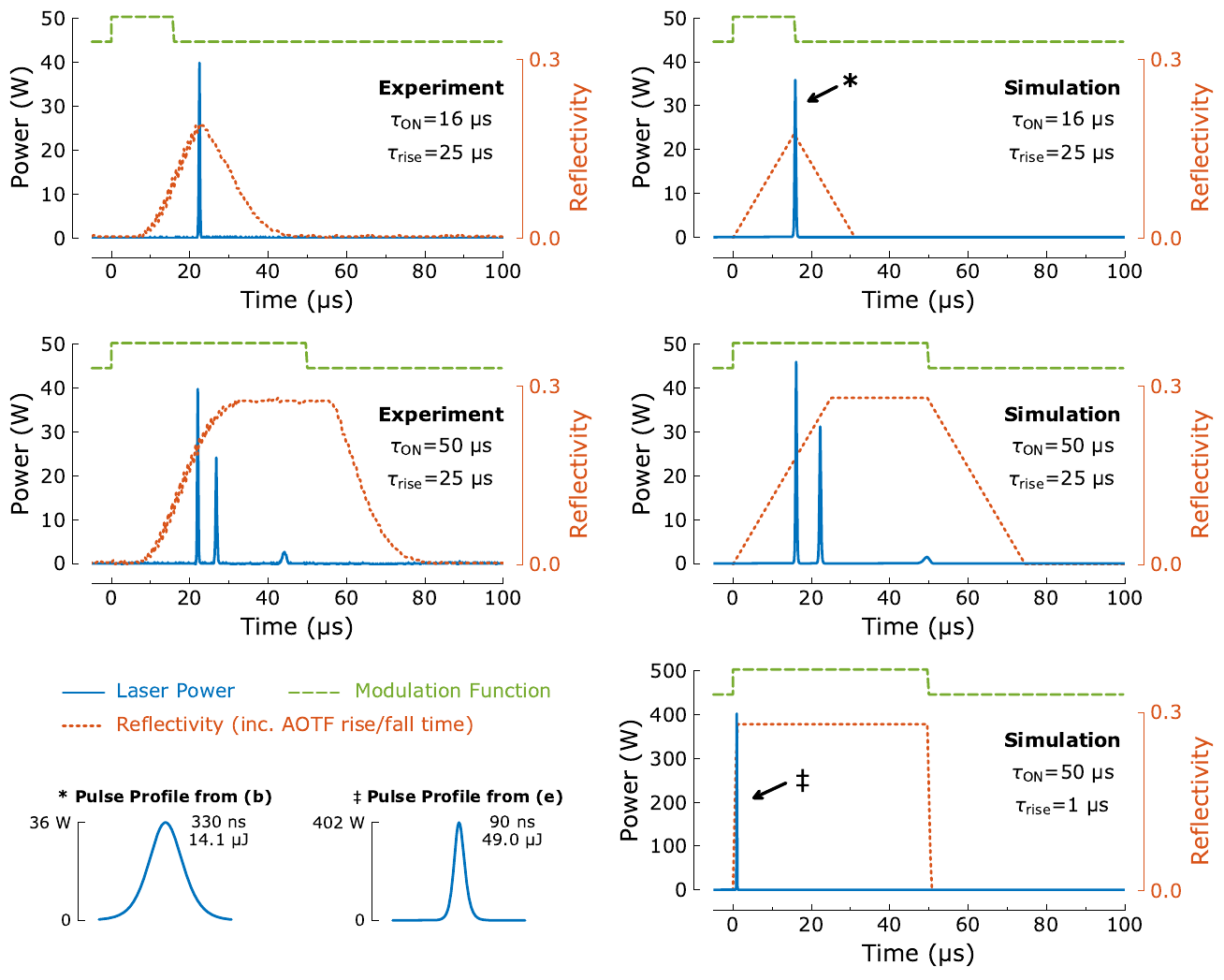}
		\put(-1, 79){ {\small (a)} }
		\put(48, 79){ {\small (b)} }
		\put(-1, 53){ {\small (c)} }
		\put(48, 53){ {\small (d)} }
		\put(48, 26){ {\small (e)} }
	\end{overpic}
	\rmfamily
	\caption{Steady-state dynamics of the actively Q-switched Dy laser at 3.1~\mum with 600~mW pump power: (a) experimental and (b) simulated single-pulse operation with optimum $\tau_\mathrm{ON}=16$~\mus AOTF on-time; (c) experimental and (d) simulated multi-pulse operation with $\tau_\mathrm{ON}=50$~\mus AOTF on-time; (e) simulated single-pulse operation with 50~\mus on-time but a faster modulator rise time of $\tau_\mathrm{rise}=1$~\mus [note the different y-axis scale compared to (a)--(d)].}
	\label{fig:sim}
\end{figure}

We implemented the model with fiber parameters matched to the experiment.
The input dichroic mirror is assumed 100\% reflective for the signal and the external cavity section incorporating the AOTF is included as a time-dependent reflectivity at the distal end of the fiber.
This time-dependent reflectivity function varies linearly from 0 to 28\% with rise time $\tau_\mathrm{rise}=25$~\mus [accounting for the measured AOTF switching time, shown in Fig.~\ref{fig:sim}(a)], where 28\% corresponds to the estimated maximum re-injected power when the AOTF is on (accounting for AOTF diffraction efficiency, Fresnel and fiber coupling losses).
As we are interested in the steady-state Q-switched laser performance, simulated pulse results are taken after at least 3~ms simulation time window, which is sufficient time for the dynamics to develop into a periodic steady state, with identical pulses generated in each modulation cycle.

Fig~\ref{fig:sim}(a) shows the experimentally measured single-pulse Q-switched output with $\tau_\mathrm{ON}=16$~\mus AOTF on-time at 600~mW pump power and 2~kHz modulation frequency. 
Simulations with these input conditions also showed stable single pulse operation [Fig~\ref{fig:sim}(b)], producing symmetrical pulses with 330~ns duration, 14.1~\muj energy and 36~W peak power.
Such properties are in reasonable agreement with the experimentally measured values of 275~ns, 11.5~\mujN, 39~W.
We note that the small disagreement could be due to uncertainties in exact glass composition between our fiber and spectroscopic data~\cite{Gomes2010} since ZBLAN is a multi-component glass with variation in ratios between manufacturers / batches.

With a longer AOTF on-time of $\tau_\mathrm{ON}=50$~\musN, the simulation also captures our experimental observation of satellite pulses being produced after the main pulse within each modulation cycle [Figs.~\ref{fig:sim}(c) and \ref{fig:sim}(d)]. 
This is an unwanted phenomena which limits the pulse peak power and could cause timing errors if used for time-resolved sensing.

Physically, this behavior can arise due to the slow switching speed (i.e. long rise time) of the modulator.
After the electrical MHz drive signal is applied to the AOTF, the cavity loss reduces over a period of $\tau_\mathrm{rise}=25$~\mus as acoustic waves build-up and diffract light towards the cavity mirror. 
A pulse will start building-up in the cavity, however, as soon as the loss falls below the amount of stored gain, which can be before the AOTF has finished switching.
In this case, the pulse build-up can be faster than the rate of switching, forming a complete pulse before the cavity loss minima is reached.
Residual inversion remains in the doped fiber, therefore, and as the loss continues to gradually fall, the emission process repeats to extract this energy: thus forming multiple low-energy pulses per Q-switching cycle~\cite{Siegman1986}.

To evaluate if this was the limiting factor in our current laser design, we repeated the simulation, but reduced the modulator rise time to $\tau_\mathrm{rise}=1$~\mus [Fig.~\ref{fig:sim}(e)].
The results show substantial improvement, predicting a 90~ns-duration pulse with 49~\muj pulse energy and 402~W peak power---an order of magnitude improvement.
This shows that with faster switching, a single pulse `sees' larger gain available during its build-up time and thus, can extract significantly more energy.
This leads to greater depletion of the population inversion, which inhibits multi-pulsing, and also means that the AOTF on-time does not change the pulse energy / duration.
In Section \ref{sec:discussion}, we discuss opportunities to practically implement this idea, using alternative acousto-optic devices.

\section{Passive Q-switching using a black phosphorus saturable absorber}
It is also possible to replace the electrical function generator and AOTF  with a passive Q switch, i.e. a saturable absorber (SA).
Such nonlinear switching devices are mature commercial products in the near-IR, primarily based on semiconductor materials (e.g. semiconductor SA mirrors, SESAMs), but are not yet widely available for the mid-IR.
Recent research results have shown that indium-based SAs can generate pulses in the 2.7--3.0~\mum region~\cite{Frerichs1996a,Li2014l}, although their prospects for longer wavelengths are limited by the indium band edge.
Therefore, intense research effort has been focused on finding alternative SA materials, such as exploring the emerging family of 2D nanomaterials~\cite{Woodward2015_as_2d,Sobon2015,Sun2016}.

A number of recent works have successfully applied nanomaterials for mid-IR saturable absorption and pulse generation, including using graphene~\cite{Wei2013, Malouf2019}, Dirac semi-metal Cd$_3$As$_2$~\cite{Zhu2017}, transition metal dichalcogenides (e.g. WS$_2$)~\cite{Wei2016} and black phosphorus (BP)~\cite{Qin2015, Qin2018}.
Of these, there is particular interest in BP for mid-IR nonlinear photonics as it is a layered material that possesses a thickness-dependent direct bandgap, varying from 0.3~eV ($\sim$4.1~\mumN) in bulk form (i.e. many layers) to 2~eV (0.62~\mumN) as a mono-layer, in addition to exhibiting ultrafast relaxation dynamics~\cite{Wang2016c, Zhang2018}.
Beyond these unique optoelectronic properties, scalable fabrication techniques for BP have also recently been developed, e.g.\ using solution processing methods, which offer great flexibility for integrating BP with existing optical components~\cite{Hu2017}.
For example, the material can be embedded in a composite polymer thin film and sandwiched between fibers~\cite{Jin2018},  coated onto side-polished fibers~\cite{Park2015a} or be directly deposited onto optical substrates / mirrors, even using highly versatile inkjet printing approaches~\cite{Hu2017}.
We therefore choose to consider BP for Q-switching in the 3.0--3.3~\mum region for the first time, which is uniquely covered by the dysprosium ion.

\subsection{Black phosphorus SA fabrication and characterization}
\begin{figure}[th!]
	\centering
	\sffamily
	\begin{overpic}{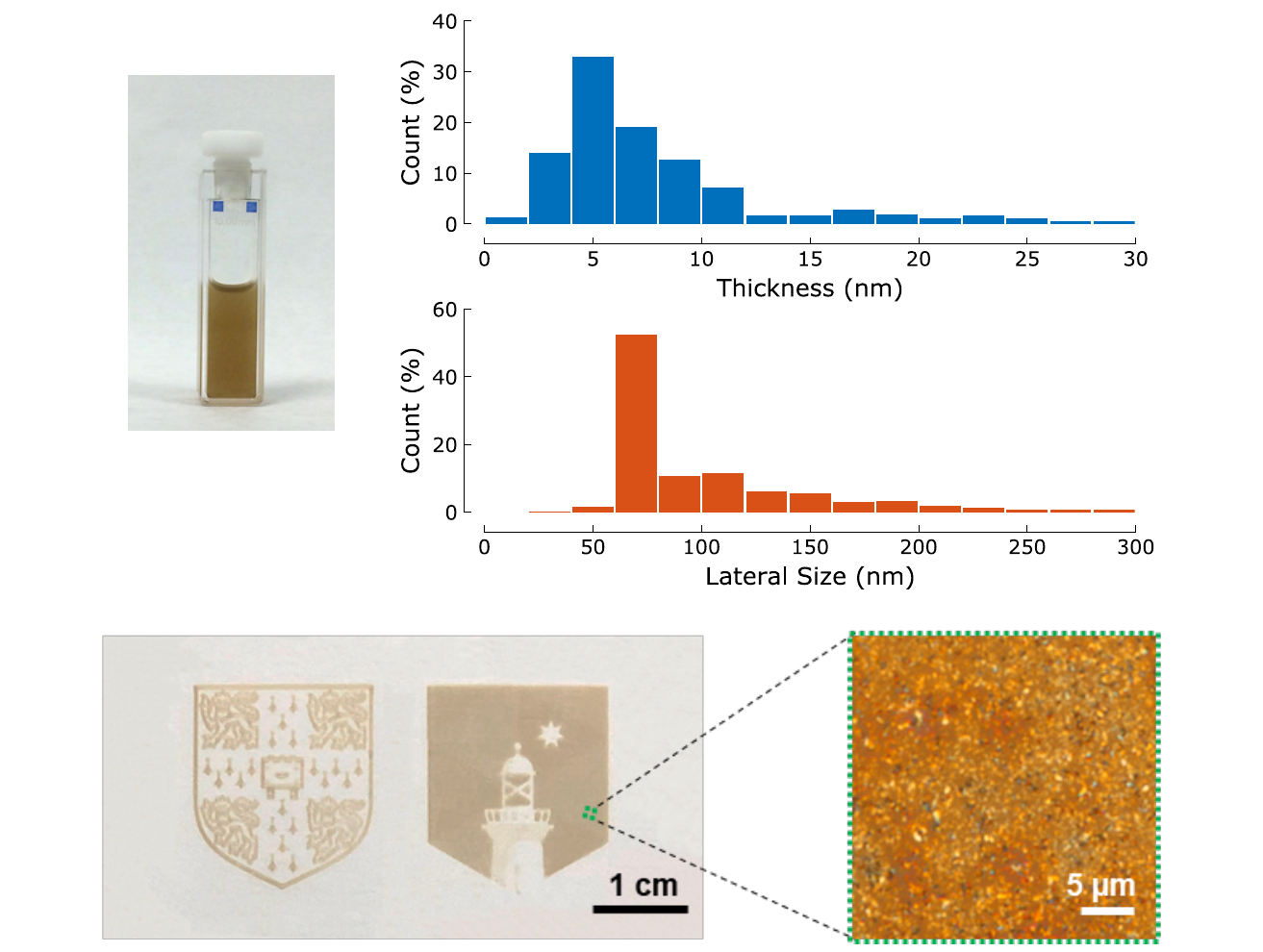}
		\put(5, 73){ {\small (a)} }
		\put(28.5, 73){ {\small (b)} }
		\put(28.5, 51){ {\small (c)} }
		\put(5, 26){ {\small (d)} }
		\put(60, 26){ {\small (e)} }
	\end{overpic}
	\rmfamily
	\caption{(a) Photograph of the as-produced black phosphorus dispersion in NMP, with a concentration of $\sim$0.5~g/L. (b) Thickness and (c) lateral size histograms of the exfoliated BP flakes from data obtained via atomic force microscopy (AFM). Averages of flake thickness and lateral size were found to be 7.8~nm and 99.8~nm, respectively. (d) Inkjet-printed BP patterns, and (e) optical microscope image corresponding to the area highlighted in (d): this exhibits continuous material deposition showing the spatial distribution of BP flakes within the pattern area.}
	\label{fig:bp_mater}
\end{figure}
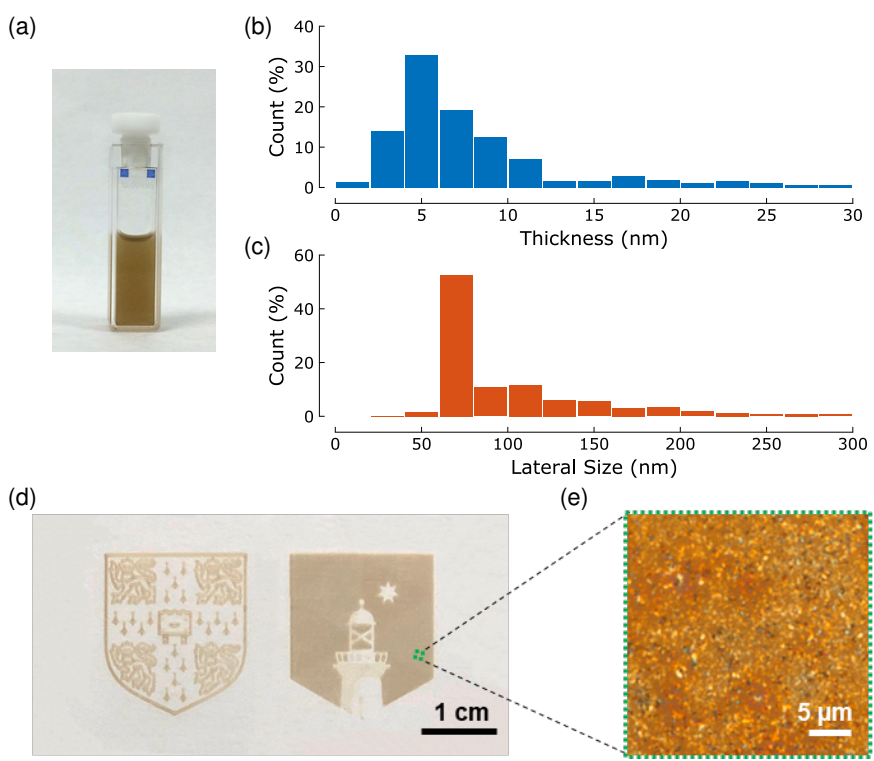

We start the SA fabrication with ultrasound assisted liquid phase exfoliation (UALPE) of BP crystals. 
UALPE is a solution processing technique that allows exfoliation of mono- and few-layer flakes from the bulk crystals in a liquid medium, under the shear forces generated by ultrasonic waves.
The liquid medium is crucial, as successful UALPE relies on an optimal intermolecular interaction; minimized enthalpy of mixing between the exfoliated flakes and the liquid. 
Previous studies show that N-methyl-2-pyrrolidone (NMP) is suited for this purpose~\cite{Hu2017,Hanlon2015}.
In this work, 10~mg BP bulk crystals (Smart Elements) are immersed in 10~mL NMP and bath-sonicated for 12 hours at 15$^\circ$C. 
To minimize oxidation of BP, anhydrous NMP is used, and the sonication tube is backfilled with nitrogen. 
The resultant dispersion is centrifuged at 1500$g$ for 30 minutes, yielding a supernatant [Fig.~\ref{fig:bp_mater}(a)] containing exfoliated BP flakes with an average thickness of $\sim$8~nm and lateral size of $\sim$100~nm [Figs.~\ref{fig:bp_mater}(b) and \ref{fig:bp_mater}(c)].

To formulate the ink, we exchange the exfoliated BP flakes from NMP into a binary solvent system of anhydrous isopropanol/2-butanol (90 vol.\%/10 vol.\%), through iterative centrifugation. 
The ink is concentrated to $\sim$5 g/L to facilitate adequate material deposition via inkjet printing. 
The formulated ink has a viscosity of $\sim$2.2~mPas, a surface tension of ~28 mNm$^{-1}$, and a density of $\sim$0.8~gcm$^{-3}$, giving an inverse Ohnesorge number of $\sim$10.
This value suggests that the ink is suitable for inkjet printing~\cite{Torrisi2012, Hutchings2012}, allowing stable generation of single droplets under each electrical pulse. 
The low surface tension of our ink also ensures adequate wetting of a wide variety of substrates including glass and polymers~\cite{Robertson2012,Santra2015,Hu2018a}.
More importantly, the binary solvent composition is shown to reduce the formation of `coffee rings' (ring like material deposition concentrated at the edges of dried droplets~\cite{Hu2018a,Deegan1997}) through the Marangoni effect: the suspended BP flakes are uniformly distributed during drying, yielding continuous material deposition.
As demonstrated in Figs.~\ref{fig:bp_mater}(d) and \ref{fig:bp_mater}(e), the ink supports high quality patterning of BP, with minimal non-uniformities. 
Such printing capability is imperative for reproducible fabrication and stable operation of inkjet-printed BP-based devices.

Having demonstrated the viability of our ink for uniform printing, we proceed to SA fabrication. 
This is done by inkjet printing of the BP ink directly onto a silver mirror. 
Following printing, the mirror is coated with a 100~nm Parylene-C layer to protect BP from degradation~\cite{Hu2017}.

\subsection{Black phosphorus Q-switched laser design and characterization}
\begin{figure}[tb]
	\centering
	\sffamily
	\includegraphics[width=\textwidth]{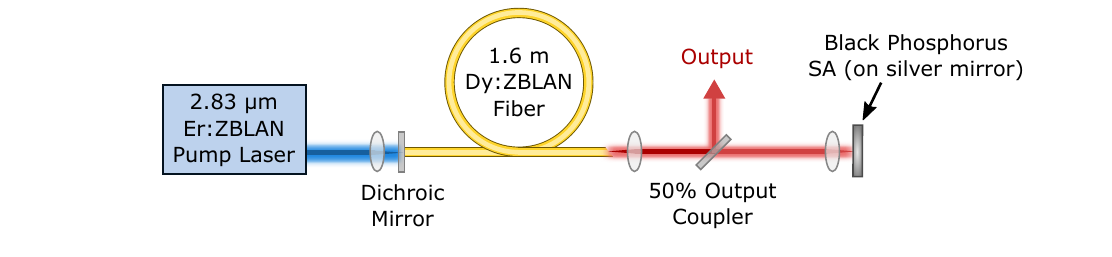}
	\begin{overpic}{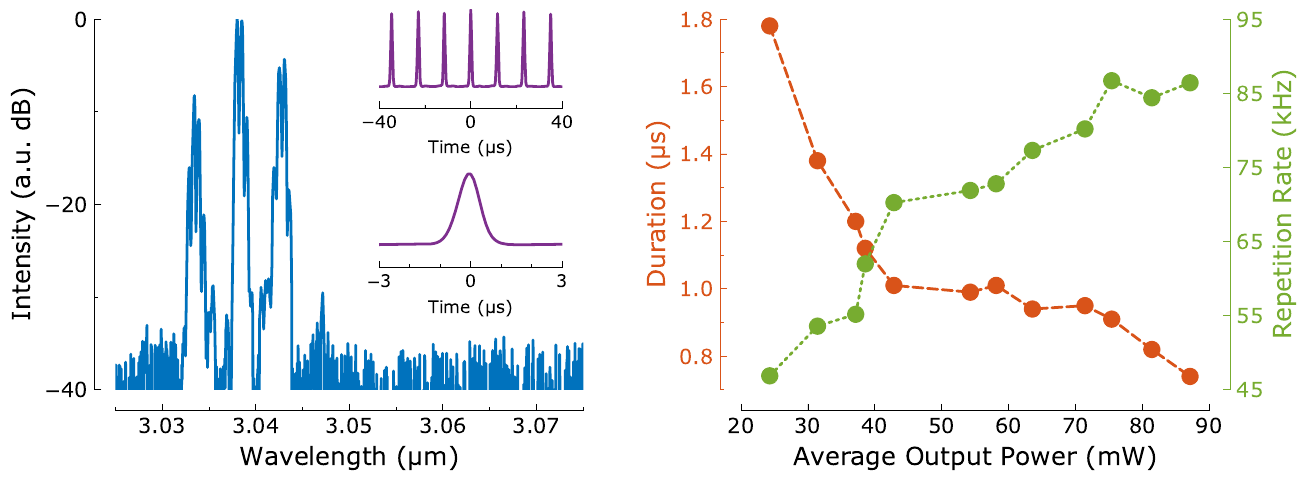}
		\put(9, 57){ {\small (a)} }
		\put(-1, 36.5){ {\small (b)} }
		\put(46, 36.5){ {\small (c)} }
	\end{overpic}
	\rmfamily
	\caption{Black phosphorus Q-switched Dy:ZBLAN laser: (a) cavity schematic; (b) optical spectrum (inset: 86~kHz pulse train \& 740~ns pulse profile); (c) variation of pulse repetition rate and duration with output power.}
	\label{fig:bp_laser}
\end{figure}

Our laser design from Section~\ref{sec:active} is adapted by replacing the AOTF with the BP-coated mirror.
An 11-mm aspheric lens is also included to focus the collimated beam to 12.5~\mum 1/$e^2$ spot diameter, and a 50\% pellicle beamsplitter is added as the output coupler [Fig.~\ref{fig:bp_laser}(a)].

For pump powers exceeding the 360~mW threshold, free-running laser emission (in the absence of an included spectral filter here) is observed, centered at 3.04~\mumN, comprising a number of closely spaced spectral peaks [Fig.~\ref{fig:bp_laser}(b)].
Temporally, the output shows a self-starting train of Q-switched pulses [Fig.~\ref{fig:bp_laser}(b) inset].

At the onset of Q-switching, pulses are generated at 47~kHz repetition rate with a full-width-at-half-maximum (FWHM) duration of 1.8~\musN, with 24~mW average output power.
With increasing pump power, the pulse duration reduces and the repetition rate increases [Fig.~\ref{fig:bp_laser}(b)], as typically observed for CW-pumped passively Q-switched lasers, related to inversion and saturation dynamics in the gain medium and SA, which are accelerated by an increased pumping rate~\cite{Siegman1986}.
The maximum achieved output power is 87~mW, for which the repetition rate was 86~kHz with 740~ns pulse duration.
This corresponds to a maximum pulse energy of 1.0~\muj and 1.3~W peak power.
At this power, long-term stable pulsation could be observed, but when the pump power increased further, the Q-switched output became unstable with noticeable amplitude fluctuation and timing jitter.
This eventually led to cessation of pulsing, which could be attributed to damage of the BP sample, as has previously been reported with other nanomaterials at high incident intensities~\cite{Wei2013}.

During stable operation, we recorded a slope efficiency of 35\%, which is the highest reported efficiency for a mid-IR Q-switched laser.
The high efficiency arises from in-band pumping of dysprosium yielding significantly lower quantum defect than typically offered by Er and Ho fiber lasers, which are pumped at 0.98 and 1.15~\mumN, respectively.
We also confirmed the role of BP in generating pulses by replacing the BP-coated mirror with a pristine silver mirror, observing only a CW output for all power levels.

\section{Discussion}
\label{sec:discussion}

\subsection{Contrasting active \& passive Q-switching}
Having investigated both active and passive Q-switches with our Dy:ZBLAN fiber laser cavity, we briefly consider the relative performance and opportunities for further enhancement.
Using the modulated AOTF, pulses were generated with durations as short as 270~ns and energies up to 12~\muj (39 W peak power), for 30~mW  average output power.
By contrast, the BP-based laser produced much longer pulses (740~ns) at lower energies (1.0~\muj for 87~mW average power).
It should be noted that other factors such as fiber length and pump laser will affect the output properties, although these were fixed in our study.
Therefore, for applications such as materials processing and pumping nonlinear fibers for supercontinuum generation, we conclude that the higher intensities offered by active Q-switching will be advantageous. 
The ability to precisely determine the wavelength using the AOTF is also beneficial, compared to the free-running multi-peaked spectra of the passively Q-switched laser.

However, the passive Q-switching geometry is significantly simpler and more compact, replacing drive electronics and the AOTF with a single coated mirror.
This is a  major advantage for practical deployment, particularly in mobile units where mid-IR LIDAR is particularly attractive.
We also note that there are promising opportunities to develop \emph{all-fiber} passively Q-switched mid-IR lasers, ruggedizing the system by replacing mirrors with fiber Bragg gratings / BP-coated side-polished fibers. 
An additional interesting future prospect is to develop all-fiber nonlinear polarization evolution (NPE)-based SAs using recently demonstrated tilted fiber gratings in fiber as polarizers~\cite{Wang2017c, Bharathan2018a}.

\subsection{Choice of acousto-optic modulator}
Our simulations revealed the slow AOTF rise time to be a limiting factor in this study, with an order of magnitude improvement (50~\mujN, 400 W peak power) predicted if the rise time could be reduced from 25~\mus to 1~\musN. 
Fortunately, mid-IR acousto-optic modulator (AOM) devices with this property are already commercially available (and have been applied to Er and Ho ZBLAN lasers~\cite{Hu2013b}). 
AOMs are distinguished from AOTFs by operating with different phase matching conditions and employing longitudinal acoustic waves (with greater acoustic velocities) rather than shear waves, thus delivering sub-microsecond switching times~\cite{Ward2018}.
The downside, however, is that AOM operation is broadband and does not include an explicit spectral filtering effect, thus removing the tunability which is an attractive feature of our laser.
AOM-based Q-switching of dysprosium lasers is thus a worthwhile topic for future study.

Finally, it should also be briefly noted that acousto-optic devices frequency shift the cavity light each round trip, due to the Doppler shift from interaction with propagating acoustic waves.
Under certain circumstances, even with constant pump power and un-modulated operation, this can disturb the steady-state inversion to produce sustained relaxation oscillations, similar to a Q-switched output~\cite{Woodward2018_fsf}.
Such pulsation is low energy, however, and not actively stabilized; thus this is not a practical Q-switching alternative.

\subsection{Fiber gain media for high-energy mid-IR pulse generation}
Finally, we note that the upper state lifetime is an important measure of gain storage potential in assessing an ion's potential for high-energy Q-switching. 
For pulse generation beyond 3~\mumN, Dy's 650~\mus lifetime~\cite{Gomes2010} compares favorably to the 3.5~\mum Er transition's value of 177~\musN~\cite{Bawden2018}, suggesting increased energy-storage potential for higher power pulse generation.
However, both these lifetimes are significantly shorter than those for transitions in Er (7.9~ms) and Ho (3.5~ms) in the 2.7--3.0~\mum window (note that all lifetimes are quoted for ZBLAN host material).
There is also potential for 4~\mum Q-switched pulse generation by considering nascent higher-energy transitions of Ho and Dy in indium fluoride glass~\cite{Majewski2018c, Maes2018a}, although further work is still needed to understand the spectroscopy and CW operation of these 4~\mum sources before optimized high-energy pulses can be achieved.

\section{Conclusion}
In summary, we have reported Q-switching of dysprosium fiber lasers for the first time: a new approach for fiber-based pulse generation beyond 3~\mumN.
Active Q-switching was demonstrated using an AOTF, achieving up to 12~\muj pulses with 270~ns duration, with the ability to tune the laser wavelength from 2.97 to 3.23~\mumN.
Q-switching dynamics were investigated, experimentally and numerically, highlighting promising opportunities for even higher pulse energies by employing an AOM rather than an AOTF.
We also fabricated an inkjet-printed black phosphorus saturable absorber as an alternative modulation technique---achieving passive Q-switching in a simpler cavity setup.
Pulses with 1.0~\muj energy and 740~ns duration were produced, demonstrating the potential of this emerging nanomaterial for nonlinear photonics in the mid-infrared region.
These advances therefore extend the parameter space of long-wavelength laser technology.

\section*{Funding}
Australian Research Council (ARC) (DP170100531).
Engineering and Physical Sciences Research Council (EPSRC) (EP/L016087/1).

\section*{Acknowledgments}
RIW acknowledges support through an MQ Research Fellowship.


\end{document}